\documentclass[aps,groupedaddress,a4paper,twocolumn,prl,showpacs]{revtex4-1}
\usepackage{graphicx}
\usepackage{amsmath}
\usepackage{amssymb}
\usepackage{verbatim}
\usepackage{color}
\usepackage{dsfont}
\usepackage{bm}
\usepackage{hyperref}
\usepackage{mathtools}
\usepackage[normalem]{ulem}

\definecolor{violet}{rgb}{0.7,0,0.5}
\definecolor{newgreen}{rgb}{0,0.6,0.0}
\definecolor{grey}{rgb}{0.4,0.4,0.4}

\usepackage{amsthm}

 \theoremstyle{definition}

 \theoremstyle{remark}

\begin{document}

\title{Majorana braiding with thermal noise}
\author{Fabio L. Pedrocchi}
\author{David P. DiVincenzo}
\affiliation{JARA Institute for Quantum Information, RWTH Aachen University, D-52056 Aachen, Germany}
\begin{abstract}
We investigate the self-correcting properties of a network of Majorana wires, in the form of a trijunction, in contact with a parity-preserving thermal environment. As opposed to the case where Majorana bound states (MBSs)  are immobile, braiding MBSs within a trijunction introduces dangerous error processes that we identify. Such errors prevent the lifetime of the memory from increasing with the size of the system. We confirm our predictions with Monte Carlo simulations. Our findings put a restriction on the degree of self-correction of this specific quantum computing architecture. 
\end{abstract}


\maketitle
\paragraph{$\!\!\!\!\!$Introduction $\!$--}$\!\!\!\!\!$Taking advantage of topological states of matter to encode and process quantum information  is viewed as  a promising route towards quantum computation.  The idea is that  the quantum information is naturally protected from  decoherence so that  quantum gates  can be performed reliably by braiding  anyonic  excitations. Recent work has focused on the realization of the topological phase as described by the Kitaev wire model \cite{Kitaev2001,LiebSchulzMattis}.  Unpaired Majorana modes are predicted in this model, and, under braiding in a network of 1D wires, these modes will behave as Ising anyons, enabling a form of topological quantum computation.   Intensive theoretical \cite{SauPRL,AliceaPRB2010,LutchynPRL2010,OregPRL2010} and experimental \cite{MourikScience2012,DengNano,DasNat,RokhinsonNatPhys,FinckPRL,ChurchillPRB} investigations of nanowire hybrid systems have been undertaken to establish the existence of these ``Majorana particles'', with the hope that a network of branched nanowires provides a suitable geometry for the moving of these particles~\cite{FuKane,Alicea2011}.

While topology does give some measure of robustness, especially in the face of static disorder \cite{Review2008, BravyiKoenig}, it has been noted that Majorana qubits are not entirely immune to decoherence.  For example, these qubits are dephased by the induced splitting of nearby Majorana states due to tunneling \cite{ZyuzinPRL}.  It is also clear that it is essential to avoid parity-changing excitations, that is, injection of individual quasiparticles from the environment, which are immediately destructive of Majorana-qubit coherence \cite{GoldsteinPRB, BudichPRB, LossRainis}.   The decoherence of Majorana qubits induced by thermal excitations above the superconducting gap has been investigated in Refs.~\onlinecite{Schmidt, Hassler}.  Additional sources of noise in a trijunction setup \cite{KlinovajaLoss} and when Majorana bound states (MBSs) are physically moved \cite{Cheng2011,Karzig, Scheurer,Karzig2015,Karzig20152} have also been recently studied. 

Starting from a microscopic model, here we perform a detailed analysis of the effect of a \emph{parity-preserving} environment on the fidelity of the topological qubit. Our main focus is the interplay between thermal errors and the adiabatic motion of MBSs, within the trijunction scheme of~\cite{Alicea2011}.
Our calculations here, using a standard treatment of a generic local bosonic environment at low temperature, show that the parity-preserving environment produces unavoidable dangerous errors arising from the motion of MBSs.  These errors put an upper limit on the lifetime of the qubits, which cannot be improved by keeping the MBSs  further apart.  The motion of MBSs  causes a spatial separation of thermally produced quasiparticle pairs, such that they act to effectively change the parity, just as parity-breaking environments do.  

\paragraph{$\!\!\!\!\!$Kitaev model--}$\!\!\!\!\!$The Kitaev wire \cite{Kitaev2001}, an archetypical example of a system supporting Ising anyons, consists of a 1D chain of hopping fermions coupled by a superconducting pairing term, 
\begin{equation}\label{eq:Hamiltonian_1}
H_{S}(\tau)=-\sum_{j=1}^{L}\mu_{j}(\tau)a_{j}^{\dagger}a_{j}-\sum_{j=1}^{L-1}(t\,a_{j}^{\dagger}a_{j+1}-\Delta\, a_{j}a_{j+1}+\text{h.c.})\,.
\end{equation}
The first term describes a site- and time-dependent chemical potential $\mu_{j}(\tau)\leqslant0$. The second and third terms describe respectively nearest-neighbor hopping with $t>0$ and superconducting pairing with $\Delta=\vert\Delta\vert e^{i\theta}$.

We first analyze on the case $\mu_{j}=0\,\forall j$ and express $H_{S}$ in terms of Majorana-mode operators $a_{j}=(\gamma_{2j-1}+i \gamma_{2j})/2$ which satisfy $\{\gamma_{j},\gamma_{k}\}=2\delta_{jk}$. For the parameters we study in this paper,  $t=\vert\Delta\vert$ and $\theta=0$, $H_S$ reduces to
\begin{equation}\label{eq:top}
H_{S}=-\Delta\sum_{j=1}^{L-1}i\gamma_{2j+1}\gamma_{2j}\,.
\end{equation}
Note that the end Majorana modes, $\gamma_{1}$ and $\gamma_{2L}$, are decoupled from the rest of the chain; thus, a  zero-energy delocalized fermionic mode exists with creation operator $d_{0}=(\gamma_{1}+i\gamma_{2L})/2$. $H_S$ can be fully diagonalized using the eigenoperators $d_{j}=(\gamma_{2j}+i\gamma_{2j+1})/2$ and written as $H_{S}=\Delta\sum_{j=1}^{L-1}(2d_{j}^{\dagger}d_{j}-1)$. 
\begin{figure}[h!]
	\centering
		\includegraphics[width=0.4\textwidth]{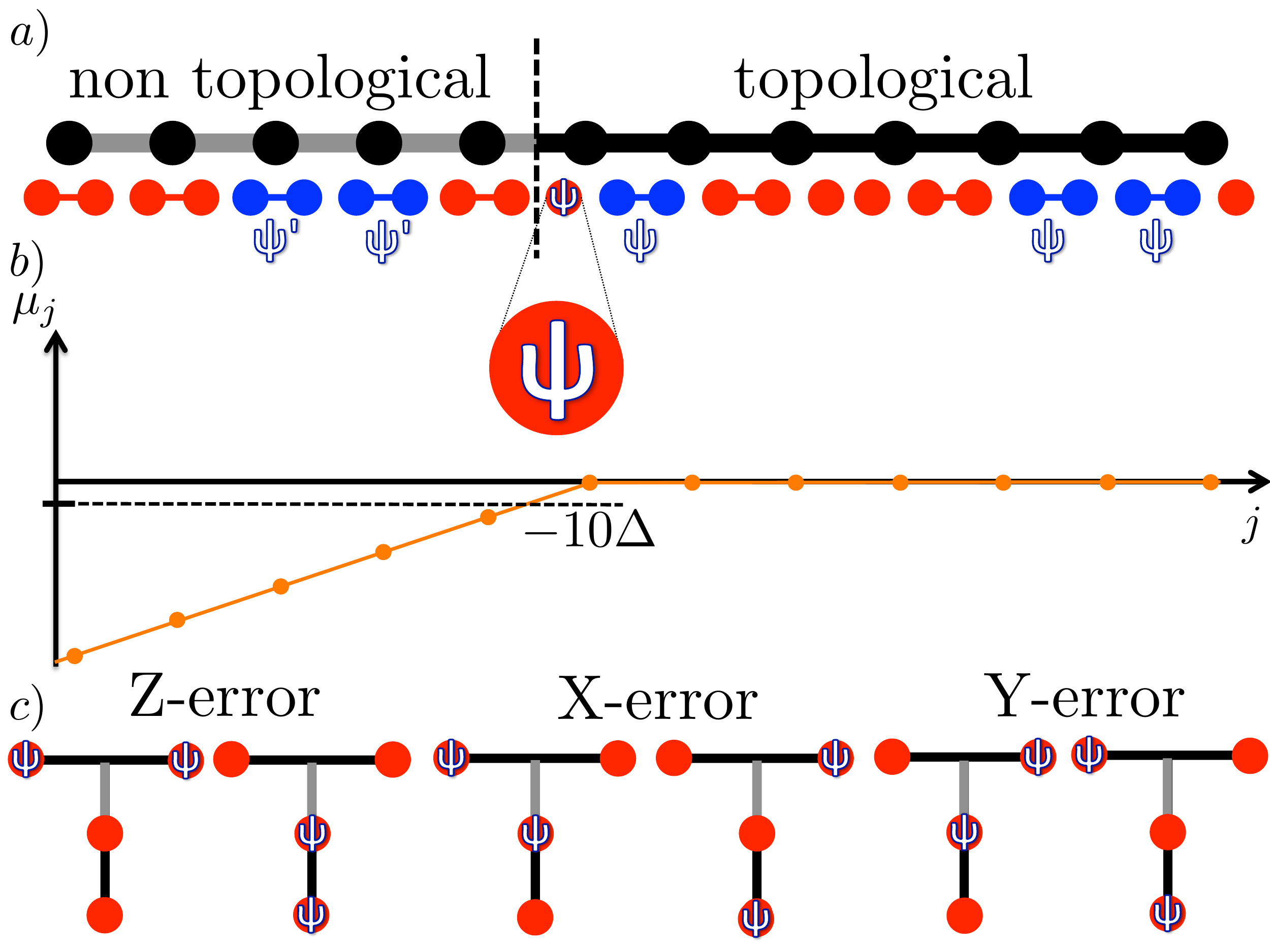}
	\caption{$a)$ Representation of the Kitaev wire supporting a nontopological (gray) and a topological (black) segment. The large black dots represent the fermionic sites of Eq.~(\ref{eq:Hamiltonian_1}). The smaller dots below represent the Majorana modes of Eq.~(\ref{eq:top}). The lines between the small dots represent the coupling between Majorana modes in the limit $\vert \mu_{j}\vert\gg t,\Delta$ (for the nontopological segment) and for the values $\mu_{j}=0$ and $t=\Delta$ (for the topological segment). The excitations above the ground states are local and can be understood as $\psi$ particles as depicted. $b)$ Space profile of the chemical potential $\mu_{j}$ corresponding to the situation in $a)$.  The chemical potential in the nontopological segment is made non-uniform  in order to localize the $\psi^{\prime}$-particles.  $c)$ Configuration of $\psi$ particles corresponding to logical $X$, $Y$, and $Z$ Pauli errors in the trijunction.}
	\label{fig:Figure2}
\end{figure}

For $\mu_{j}\ne 0$, MBSs localized near the end of the wire persist as long as $\vert\mu_{j}\vert\leqslant 2t$; this is called the \emph{topological} phase. For $\vert\mu_{j}\vert\geqslant 2t$ the localized modes disappear and we enter the \emph{nontopological}  phase. Deep in the nontopological phase, with $\vert\mu_{j}\vert\gg t,\Delta$, the Majorana Hamiltonian approaches $H_{S}^{\text{nontop}}=-i\mu/2\sum_{j=1}^{L}\gamma_{2j-1}\gamma_{2j}$ and the Majorana modes are paired in a shifted way as compared to Eq.~(\ref{eq:top}), see Fig.~\ref{fig:Figure2}a.

\begin{figure}[h!]
	\centering
		\includegraphics[width=0.5\textwidth]{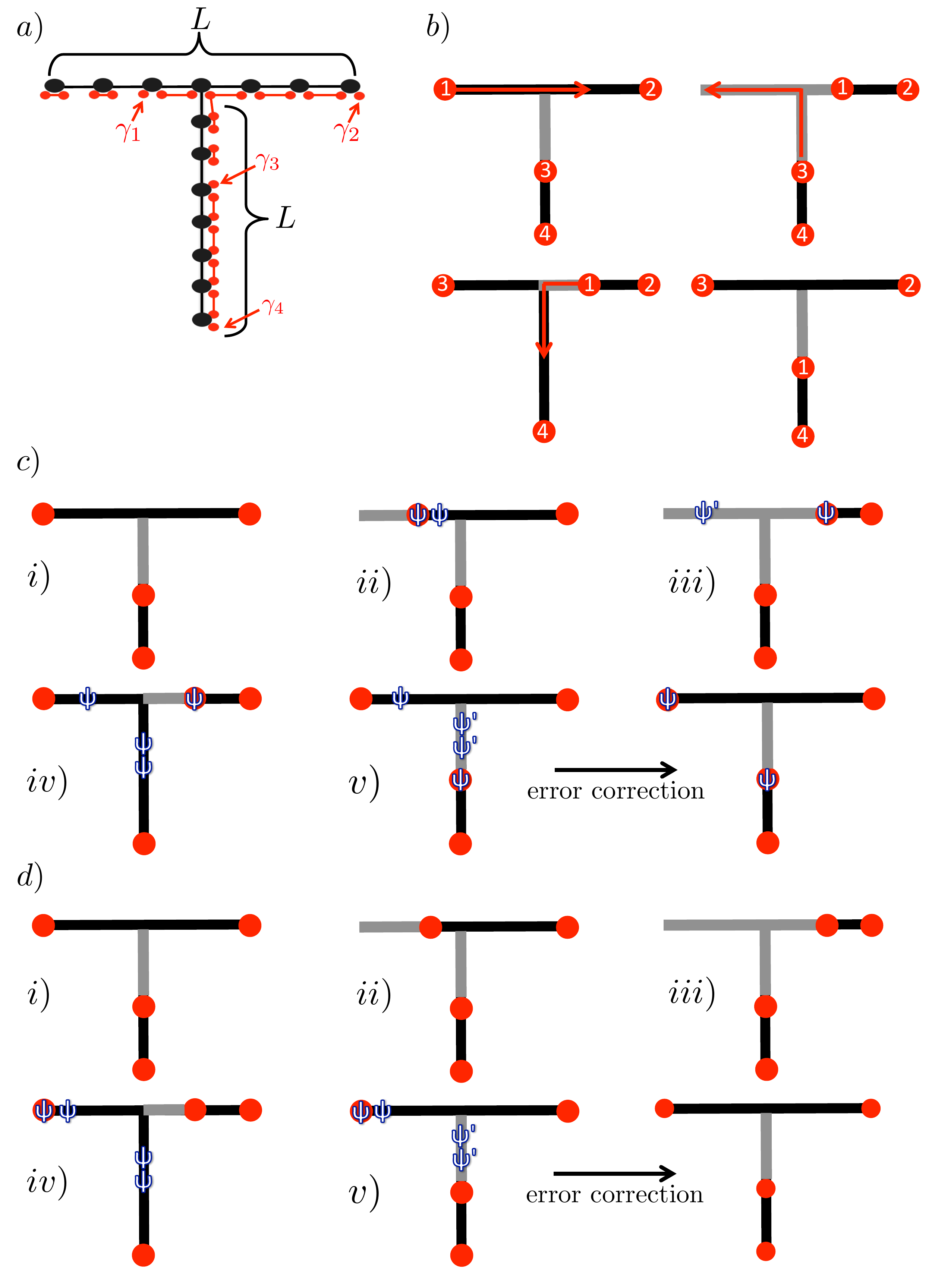}
	\caption{$a)$ Trijunction of size $L=7$, composed of horizontal and vertical wires coupled by hopping and superconducting pairing. The zero-energy MBSs  $\gamma_{1,2,3,4}$ are shown.
$b)$ Braiding sequence considered in this work; MBSs  $1$ and $3$ are exchanged, while MBSs  $2$ and $4$ remain immobile. $c)$, $d)$: Sequence of two equiprobable error processes. The two sequences lead to the exact same error syndrome. After application of our error correcting algorithm to the evolution at $v)$,  the outcome in $c)$ is faulty while the one in $d)$ is successful.}
	\label{fig:Figure1_2}
\end{figure}

More generally, MBSs occur at the junction between topological and nontopological segments of the wire. By varying the chemical potential, one can increase or decrease the size of the nontopological segments and thus move the position of the MBSs. As proposed in Ref.~\cite{Alicea2011}, this technique can be used to braid MBSs in a nanowire trijunction setup (Fig.~\ref{fig:Figure1_2}) and perform gates in a topologically protected fashion. In order to stay within the ground-state subspace, the motion of the MBSs must be adiabatic. This means that the chemical potentials $\mu_{j}(\tau)$ must be varied on a timescale  slow compared to $1/\Delta$. Specifically, here we vary the chemical potentials linearly with time. 

\paragraph{$\!\!\!\!\!$Encoding $/$ Error Correction $\!$--}$\!\!\!\!\!$
For the sake of clarity, we first focus on the encoding of a logical qubit in a single Kitaev wire with $\mu_{j}=0$ and then we generalize the discussion to the branched wire forming a trijunction. The ground-state subspace of $H_{S}$ in Eq.~(\ref{eq:top}) forms a stabilizer code with stabilizer operators $S_{j}=i\gamma_{2j+1}\gamma_{2j}$ \cite{Terhal,BravyiKoenig};  that is, the logical qubit states $\vert \bar{0}\rangle$ and $\vert\bar{1}\rangle$  are invariant under the action of the $S_{j}$ operators with $j=1,\ldots,L-1$. Furthermore, they are defined by the phase they acquire under the action of the parity operator $S_{0}=i\gamma_{1}\gamma_{2L}$, namely $S_{0}\vert\bar{0}\rangle=\vert\bar{0}\rangle$ and $S_{0}\vert\bar{1}\rangle=-\vert\bar{1}\rangle$.

In this context, excitations above the ground states are interpreted as particles; whenever a stabilizer is violated, then we say that the wire supports a $\psi$ particle at the position of the violated stabilizer, see Fig.~\ref{fig:Figure2}a. Also here we represent a parity flip $S_{0}\rightarrow -S_{0}$ by drawing a $\psi$ particle \emph{inside} one of the MBS.  In the ground-state subspace, drawing the $\psi$ inside the MBS $\gamma_{1}$  is interpreted as a logical $X$ flip of the qubit state, drawing it inside the MBS $\gamma_{2L}$  corresponds to a logical $Y$ flip, while drawing it inside both of them corresponds to a logical phase flip $XY\propto Z=S_{0}$. Note that here we consider solely parity-conserving perturbations and $\psi$ particles are thus created \emph{in pairs}. A string operator creating pairs of $\psi$'s corresponding to, say, $S_{j}=-1$ and $S_{k}=-1$ is $\mathcal{S}_{jk}=\gamma_{2j+1}\gamma_{2j+2}\cdots\gamma_{2k}$. String operators give us a rigorous way to define fusion of the different excitations, and thus perform error correction; two $\psi$ excitations $S_{a}=-1$ and $S_{b}=-1$ are fused and annihilated by applying $\mathcal{S}_{ab}$. If one wants to fuse a $\psi$ excitation $S_{a}=-1$ with the MBS  $\gamma_{1}$, one should apply $\mathcal{S}_{0a}$; this moves the $\psi$ from the bulk to inside the MBS,  thereby flipping the parity $S_{0}\rightarrow -S_{0}$. 
This phenomenology corresponds to the standard fusion rules for Ising anyons \cite{BondersonThesis} with topological charges $\sigma$, $\psi$, and $1$: $\sigma\times\sigma=1+\psi$, $\psi\times\psi=1$, and $\sigma\times\psi=\sigma$. MBSs  are identified with the Ising anyons $\sigma$, because they follow the prescribed fusion rules (as just shown) as well as braiding rules \cite{Alicea2011}. 

We now discuss the trijunction setup. We consider a scenario with four MBSs $\gamma_{1-4}$, using a scheme in which $\gamma_{1}$ and $\gamma_{3}$ are braided while $\gamma_{2}$ and $\gamma_{4}$ remain immobile, see Fig.~\ref{fig:Figure1_2}b.  In such a case, the ground-state subspace is 4-fold degenerate and we follow the procedure presented in Ref.~\cite{BravyiPRA} to encode the logical qubit in a fixed-parity sector, say $i\gamma_{1}\gamma_{2}\,i\gamma_{3}\gamma_{4}=+1$. Similar to above, we have $i\gamma_{1}\gamma_{2}\vert\bar{0}\rangle=i\gamma_{3}\gamma_{4}\vert\bar{0}\rangle=\vert\bar{0}\rangle$ and $i\gamma_{1}\gamma_{2}\vert\bar{1}\rangle=i\gamma_{3}\gamma_{4}\vert\bar{1}\rangle=-\vert\bar{1}\rangle$. Again, excitations are localized and can be understood in terms of $\psi$ particles. Logical $X$-, $Y$-, and $Z$-errors are present when $\psi$ particles reside inside MBSs,  see Fig.~\ref{fig:Figure2}c.

Pairs of $\psi$ particles can be created, annihilated or can hop due to interaction with a bath. But we presume the possibility of final error correction that undoes the effect of the thermal environment and protects the stored quantum information. We thus assume that it is possible to identify the position of the $\psi$ particles in the bulk of the trijunction setup and thus identify the {\em error syndrome}. Note that the fusion rule $\sigma\times\psi=\sigma$ indicates that a $\psi$ inside the MBS is invisible to error correction. Having the syndrome information in hand, one performs a set of unitary operations having the effect of annihilating $\psi$ particles in order to map back into the ground-state subspace and hopefully recover the stored qubit. The simple algorithm we use is an adaptation of the algorithm presented in Ref.~\cite{Wootton,long}. The main idea is to fuse pairs of nearby particles (MBS or $\psi$). 

\paragraph{$\!\!\!\!\!$Coupling to a Thermal Bath--}$\!\!\!\!\!$
We see that, over time, the fundamental $\psi$ excitations cause the destruction of the encoded information. Here we derive a master equation for the trijunction when such excitations result from coupling to a thermal bath. We start with the following microscopic model:
\begin{equation}\label{eq:Htot}
H(\tau)=H^{\text{trij}}_{S}(\tau)+H_{B}+H_{SB}\,.
\end{equation}
Here $H_{S}^{\text{trij}}(\tau)$ is the adiabatically varied trijunction Hamiltonian, consisting of hopping and pairing terms as in Eq. (1) for each segment of the trijunction, $H_{B}$ 
is a bosonic bath Hamiltonian and $H_{SB}=-\sum_{j}B_{j}\otimes (2a_{j}^{\dagger}a_{j}-1)=-i\sum_{j}B_{j}\otimes\gamma_{2j-1}\gamma_{2j}$  is a conventional system-bath coupling. Importantly, $H_{SB}$ conserves the overall parity of the system; system operators $\gamma_{2j-1}\gamma_{2j}$ anticommute with two stabilizers $S_{j-1}$ and $S_{j}$, thus $\psi$ excitations are created in pairs. 

Following the Davies prescription \cite{long}, we obtain the master equation, in Lindblad form, for the system density matrix in the adiabatic limit,
\begin{equation}
\dot{\rho}_{S}(\tau)=-i\,[H^{\text{trij}}_{S}(\tau),\rho_{S}(\tau)]+\mathcal{D}(\rho_{S}(\tau))\,,
\end{equation}
with
\begin{eqnarray}
\mathcal{D}(\rho_{S}(\tau))&=&\sum_{i,j}\sum_{\omega(\tau)}\gamma^{ij}(\omega(\tau))\left(A^{i}(\omega(\tau))\rho_{S}(\tau){A^{j}(\omega(\tau))}^{\dagger})\right.\nonumber\\
&&\left.\hspace{1cm}-\frac{1}{2}\{{A^{j}(\omega(\tau))}^{\dagger}A^{i}(\omega(\tau)),\rho_{S}(\tau)\}\right)\,,
\end{eqnarray}
where $\omega(\tau)$ describes the different energy gaps at time $\tau$ between the different eigenbranches of $H_{S}^{\text{trij}}(\tau)$. The Davies approach gives jump operators $A^{i}(\omega(\tau))$ satisfying detailed balance, that cause transitions between different system eigenbranches separated by energy $\omega(\tau)$ at time $\tau$.  A crucial property of our system Hamiltonian is that the jump operators remain strictly local \cite{long}.

The diagonal matrix elements of $\rho_{S}(\tau)$ decouple from the off-diagonal elements \cite{long} so that one obtains a Pauli master equation for the population in an instantaneous eigenstate $\vert n(\tau)\rangle$ of $H_{S}^{\text{trij}}(\tau)$,
\begin{equation}\label{eq:Pauli}
\frac{dP(n(\tau),\tau)}{d\tau}=\sum_{m(\tau)}W(n(\tau)\vert m(\tau))P(m(\tau),\tau)-m\leftrightarrow n\,,
\end{equation}
$$W(n(\tau)\vert m(\tau))=\gamma(\omega_{mn}(\tau))\vert\langle m(\tau)\vert A^{i_{mn}}(\omega_{mn}(\tau))\vert n(\tau)\rangle\vert^{2}$$ and $\gamma(\omega)=\kappa\left\vert \omega/(1-\exp(-\beta\omega))\right\vert$ is the bath spectral function with inverse temperature $\beta$ and with coupling constant $\kappa$. Note that $\gamma(\omega)$ does not depend on position since we assume the baths coupled to each site to have identical form and thus to have the same spectral function.

For the sake of illustration, we first consider here the time-independent case, i.e. when the MBSs are immobile, and identify which transitions are caused by the jump operators. In the case of moving MBSs many more fundamental processes are allowed, but we believe that the following list gives some intuition for the error processes that also occur in the more general case. i) Creation or annihilation of a $\psi$-pair in the bulk, with energy cost $\pm4\Delta$. ii) Hopping of a $\psi$ to a nearest neighbor site in the bulk, with energy cost $0$. iii) $\psi$-pair creation (annihilation) at a boundary supporting a MBS,  with energy cost $\pm 2\Delta$. iv) Hopping of a $\psi$ into or out of a neighboring MBS,  with energy cost $\pm2\Delta$. In Ref.~\cite{long}, we present an exhaustive list of the hundreds of allowed error processes 
that occur in the trijunction during adiabatic braiding.

It is worth pointing out that the system-bath interaction in Eq.~(\ref{eq:Htot}) does not support the creation of $\psi$ particles in the nontopological segments of the trijunction.  Indeed, as the chemical potential becomes very negative, the eigenstates become very close to those of $H_{S}^{\text{nontop}}$ and  $[H_{S}^{\text{nontop}},H_{SB}]=0$.  However, this does not mean that no $\psi$ particles will ever be present in the nontopological segments. To understand that, one needs to investigate the interaction between moving MBSs  and $\psi$ particles. We note: i) When a MBS  moves over an existing $\psi$ particle, the $\psi$ particle is transferred from the topological segment into the nontopological one. For clarity we use the notation $\psi^{\prime}$ for an excitation in the nontopological segments.  In the reverse scenario, when a MBS  moves over a $\psi^{\prime}$, then the excitation transfers back to the topological segment, becoming a $\psi$. ii) A $\psi$ particle inside a MBS  remains inside the MBS  during the motion. These two properties can be proved straightforwardly by diagonalization of a four-site model \cite{long}.  In our error-correcting algorithm presented above, $\psi$ and $\psi^{\prime}$ particles are treated in the exactly same way. In particular, during error correction it is assumed that the position of excitations can be measured everywhere, not only in the topological segments.

\paragraph{$\!\!\!\!\!$Dangerous Error Processes--}$\!\!\!\!\!$
We show that some elementary error processes lead to a lifetime of the stored logical qubit that cannot be improved by increasing the size of the trijunction setup. This is in contrast with the case of immobile MBSs,  where the lifetime grows with the system size \cite{BravyiKoenig, long}; in Ref.~\cite{long} we show that the lifetime grows logarithmically with $L$ in the high-temperature limit. Immobile MBSs thus represent a good quantum memory. However, a logarithmic scaling of the lifetime is generally too weak if braiding is performed since the time to adiabatically braid the MBSs scales linearly with $L$ \cite{Beverland}. Here we show that the lifetime does not actually scale at all with $L$ when MBSs are moved because of dangerous error processes. These error processes thus put a restriction on the degree of self-correction of this specific quantum computing architecture when braiding is executed.  
\begin{figure}[h!]
	\centering
		\includegraphics[width=0.4\textwidth]{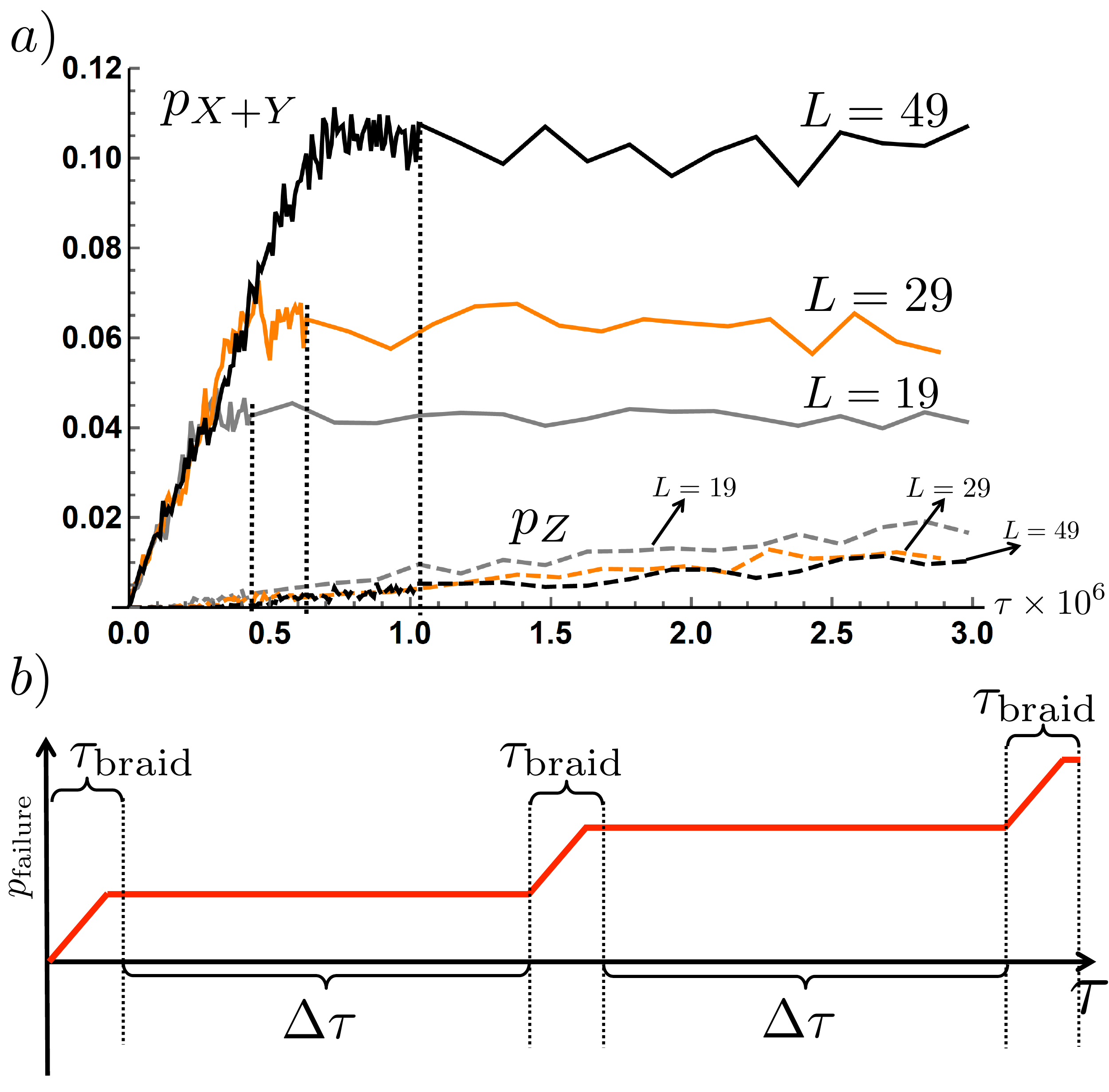}
	\caption{$a)$ Probabilities $p_{X+Y}$ (solid) and $p_{Z}$ (dashed) as a function of time for trijunctions with $L=19,29,49$. The elementary time step to move a MBS  to a nearest-neighbor site is $10^{4}/\Delta$. The braiding time is $\tau_{\text{braid}}=(2L+2)10^{4}/\Delta$; a dashed vertical line specifies when $\tau_{\text{braid}}$ is achieved. For times $\tau<\tau_{\text{braid}}$, we execute braiding motion incompletely while the coupling to the external bath is on. We then finish the braiding motion unitarily without any coupling to the thermal bath. $b)$ Artistic representation of the total probability of failure as a function of time for a trijunction setup of a given length where three braiding motions, separated by a time interval $\delta \tau$, are executed. If the coupling to the external bath is small, the only contribution to $p_{\text{failure}}$ is due to dangerous errors during braiding.}
	\label{fig:Figure3}
\end{figure}

Consider the creation of a $\psi$-pair, one inside a MBS  and one in the bulk of the trijunction. Since a $\psi$ inside a MBS  cannot escape when the MBS  is moved, this $\psi$ particle will be dragged along during the braiding motion. In other words, an originally local error process becomes highly nonlocal due to the braiding motion. For example, the error sequence depicted in Fig.~\ref{fig:Figure1_2}c leads to a logical $X$ error after our error correcting algorithm has been applied, although only two fundamental error processes occurred. Dangerous error processes will generally lead to $X$ and $Y$ logical errors and our error correcting algorithm will fail independent of the system size. It may appear surprising that $X$ and $Y$ errors are at all possible since the total parity of the system is conserved. However, this can happen because the parity of individual topological segments is not conserved since MBSs  are moved and $\psi$ particles can be transferred to nontopological segments as mentioned above.

One natural question is whether a better algorithm would be able to take into account the non-locality introduced by braiding. Unfortunately, this is impossible, as shown by Figs.~\ref{fig:Figure1_2}c and d. Here we see two distinct sequences of two fundamental error processes leading to exactly the same error syndrome. The difference between the two final states is the occurrence of the $\psi$ particles inside MBSs. But these $\psi$'s are invisible to error correction, so these two situations produce the same error syndrome and are not distinguishable by {\em any} error correcting algorithm. If one of these cases is successfully recovered by error correction, the other will lead to failure. Our discussion is independent of the size of the trijunction; this implies that the lifetime of the stored information will not grow with system size.
Our study thus provides evidence that active error correction \emph{during braiding} would be necessary in topological quantum computation \cite{WoottonLoss, BrellPRX, Hutter, WoottonHutter}. 
\paragraph{$\!\!\!\!\!$Monte Carlo--}$\!\!\!\!\!$
We confirm our predictions by simulating the Pauli master equation (\ref{eq:Pauli}) with standard Monte Carlo methods. We have calculated the rates of all the allowed error processes by numerically diagonalizing six-site trijunctions, see Ref.~\cite{long} for more details. We focus on the low-temperature regime $\beta=4/\Delta$.  A discussion of the high-temperature case can be found in Ref.~\cite{long}.

Fig.~\ref{fig:Figure3}$a$ shows the probabilities $p_{X+Y}$, and $p_{Z}$ of $X$ and $Y$, as well as $Z$ logical errors as a function of time for trijunctions with different $L$ \cite{footxyz}. A single braid (Fig. 2b) is executed at the beginning and the MBSs  are then left immobile for the remaining time. $p_{X+Y}$ increases significantly during the braiding period. This is in agreement with the above discussion of dangerous processes that occur solely during the motion of  MBSs.  Also, the growth of $p_{X+Y}$ at small times is found to be independent of $L$, confirming that dangerous processes lead to a lifetime independent of the system size. On the other hand, $Z$ logical errors are due either to a high density of $\psi$ particles or to diffusion of a $\psi$ particle over a distance greater than $L/2$. In both cases error correction will eventually fail due to an accumulation of error processes. Therefore, $p_{Z}$ stays small during the (short) braiding period and increases at larger times.  

\paragraph{$\!\!\!\!\!$Conclusion--}$\!\!\!\!\!$ 
While our results put restrictions on the
quantum coherence that may be expected in one particular braiding
scheme, there are many other scenarios in which MBSs  could
have unique strengths in the processing of quantum information.  For
example, the present analysis does not restrict the feasibility of the
interaction-induced braiding concept of Refs.~\cite{Hassler2012, Fulga}; we hope
to pursue an extension of our system-environment treatment to this case. 

\paragraph{$\!\!\!\!\!$Acknowledgements--}$\!\!\!\!\!$We are happy to thank Nick Bonesteel, Stefano Chesi, Fabian Hassler, Adrian Hutter, and Daniel Loss for valuable discussions, and we thank the Alexander von Humboldt foundation for support.


\begin{thebibliography}{100}
\bibitem{footxyz} It is worth pointing out that the effective error model that applies to the logical qubit after error correction has been implemented is a generalized depolarizing channel, $\rho_{S}\rightarrow \sum_{O=I,X,Y,Z}p_{O} \,O\,\rho_{S} \,O^{\dagger}$ , see Ref.~\cite{long} for more details.
\end{thebibliography}
\end{document}